# On the SOA-based MZI all-optical logic gates for all-optical networks

Qing Zheng
Department of Computer Sciences
China University of Mining and Technology
Beijing, China
Email: qzheng@ome.ac.cn

**Abstract**

In this paper, an all-optical logic scheme which exploits the cross-phase modulation (XPM) effect in semiconductor-optical-amplifier-assisted Mach-Zehnder Interferometer (SOA-MZI), is proposed, performance analyzed and parameters optimized. The proposal is validated and the system performance under various parameters is examined through numerical simulations. With only moderate parameters, high-speed all-optical AND gate based on SOA-MZI is realized with fairly high performance. The results are helpful for designing of SOA-based all-optical logic devices.

## 1. Introduction

All-optical logic devices are vital elements in ultra-high speed all-optical networks. Particularly, optical Boolean AND operation is indispensable to critical networking functions, such as header recognition, self-routing, switching, signal regeneration, and data encoding and encryption.

Various schemes of all-optical AND operation have been reported, such as 10Gb/s AND using electro-absorption modulators (EAM) [1], 2.5Gb/s AND using randomly birefringent fiber (BRF) in a nonlinear optical loop mirror (NOLM) [2], 10Gb/s AND in a nonlinear DFB waveguide monolithically integrated with a Y-coupler [3]; AND exploiting the effect of four-wave mixing (FWM) [4] or cross-polarization modulation (CPM) [5] in a SOA; AND based on SOA-MZI [6], gap soliton formation in a fiber Bragg grating [7], soliton trapping in a BRF followed by a frequency filter [8], a PNPN-type optoelectronic integrated functional device (OFD) [9], or ultrafast carrier dynamics in a multi-quantum-well semiconductor optical amplifier [10].

Given that SOA-MZI-based devices possess the practical advantages of low power consumption, low latency, high stability and integration potential, devices of this type have become the subject of much intensive research in many laboratories [11][12]. In this paper, a novel scheme of high-speed all-optical AND gate exploiting the XPM effect in SOA-MZI is proposed and investigated theoretically. In case of various system parameters, performance of the AND gate is assessed and thereby advice on key parameter designing is given. Besides, the data-pattern dependence of the gate and the speed limit are discussed.

## 2. Operation principle

The optical AND gate in our study consists of a symmetrical MZI with one SOA located in the same relative position of each arm, as shown in Fig. 1. For the Boolean calculation $A \cdot B = C$, the input logical signal $A$ enters Port 1 and splits via Coupler C1 into two equal parts, named after the clockwise (CW) one and the counter-clockwise (CCW) one respectively, acting as the probe signals. The input logic signal $B$ is amplified and divided into two equivalent parts, $B_1$ and $B_2$, functioning as the control signals. $B_1$ is introduced into the upper arm via Wavelength Multiplexer WDM1, and $B_2$ into the lower arm via WDM2 after a time delay $\Delta\tau$.

In Fig. 1, the control and the probe pulses counter-propagate along both the arms, it is called colliding pulse MZI (CMZI). In the case of the control and the probe pulses co-propagating, we name it traveling pulses MZI (TMZI). Particularly, we pay more attention to CMZI in this study, for which a filter is unnecessary at Port 3 to reject the control signals and thus the device is more compact.

If the control signal $B$ is ZERO, the SOA-MZI is balanced and no signal emerges from Port 3. In contrast, if $B$ is ONE, due to the XPM effect in the SOAs, a differential phase-shift is introduced briefly to the probe signals in both arms so as to switch them out of Port 3. Thus, signal $C$ at Port 3 is actually the result of Boolean calculation $A \cdot B$.

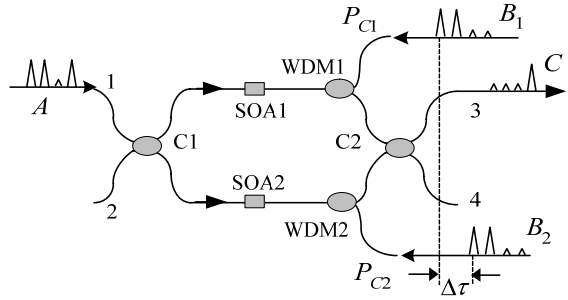

Fig. 1 Configuration of all optical AND gate using SOA-MZI

In configuration, signal $A$ and $B$ should be synchronized in advance. Moreover, the SOAs are expected to be identical and the two arms of MZI are of the same length. If there is polarization effect, switching will happen even when either $A$ or $B$ is ZERO. Thereby, both SOAs should be polarization independent. Otherwise, the polarizations of both control signals should be the same and are maintained as they traverse the arms.

## 3. Theoretical Analysis

For simplicity, we neglect the loss and the amplified spontaneous emission (ASE) noise in SOA. Also, we neglect the group velocity dispersion, since its effect on the pulse propagation is negligible along the SOA length (several hundreds of micrometers). Thus the clock and the control signals travel through the SOA at the same speed.

Thereafter, we define the pulse transmission time in the SOAs as $T_{TRAN} = L n_{SOA}/c$, where $n_{SOA}$ is the effective index of the SOAs. When $T_{TRAN}$ is comparable to the pulse-width, as in this paper, the SOA length ($L$) and the nonlinear gain compression effect must be taken into consideration.

The generic gain and phase dynamic response of SOA under picosecond control pulses has already been studied in [*8]. More accurate theoretical models have been discussed in [*9] and [*10].

However, since the input pulse energy is less than 1pJ in this paper, we treat the SOAs with a simplified model by J. M. Tang, which does not include the effects of two-photon absorption (TPA) and high-speed nonlinear refraction (UNR), as follows [*11].

$$\frac{\partial g(z,t)}{\partial t} = \frac{g_0 - g(z,t)}{\tau_C} - \frac{g(z,t)}{1+\varepsilon P_C(z,t)} \frac{P_C(z,t)}{E_{SAT}} \quad (1)$$

$$\frac{\partial P_C(z,t)}{\partial z} = \frac{g(z,t)}{1+\varepsilon P_C(z,t)} P_C(z,t) \quad (2)$$

$$\frac{\partial \phi_C(z,t)}{\partial z} = -\frac{1}{2}\alpha \frac{g(z,t)}{1+\varepsilon P_C(z,t)} \quad (3)$$

where $P_C(z,t)$ and $\phi_C(z,t)$ are the power and the phase-shift of the control signal. $g_0$ is the small signal gain. $\tau_C$ is the spontaneous carrier lifetime. $\alpha$ is the linewidth enhancement factor. $\varepsilon$ is the nonlinear gain compression factor due to the effect of carrier heating and spectral hole-burning. $g(z,t)$ corresponds to the instant gain coefficient at a place ($z$) of the SOA. $E_{SAT}$ is the saturation energy and satisfies.

$$E_{SAT} = P_{SAT} \cdot \tau_C = \frac{\hbar \omega_0 \tau_C}{a_N} \cdot \frac{dW}{\Gamma} \quad (4)$$

where $h$ is the Plank constant, $\omega_0$ is the control pulse center wavelength, $a_N$ is the general gain factor, $d$ and $W$ are the depth and the width of the SOA active region, and $\Gamma$ is the general confinement factor.

To modeling the characteristics for both co- and counter- propagation of the control signals with respect to the probe clock, we design the probe clock intensity so small as not to modify the optical properties of the SOAs, and take the transparency assumption similar to the approaches in [12] and [13]. Namely, the SOA is quasi-transparent to both the control pulse and the gain coefficient. The instant gain coefficients ($h$) is then calculated by

$$h(t) = \int_{-\infty}^{\infty} d\tau \int_{-L/2}^{L/2} g(z,\tau)\delta(\tau - t \pm zn_{SOA}/c)dz \quad (5)$$

The integral SOA gains ($G_1$ and $G_2$) and the pulse phase-shifts ($\Delta\phi_1$ and $\Delta\phi_2$) in upper and lower arms are expressed as

$$G_1(t) = \exp[h(t)], \quad G_2(t) = \exp[h(t-\Delta\tau)] \quad (6)$$

$$\Delta\phi_1(t) = -\alpha h(t)/2, \quad \Delta\phi_2(t) = -\alpha h(t-\Delta\tau)/2 \quad (7)$$

In Eq. (5), the operators "+" and "−" correspond to counter- and co- propagation respectively. To solve the equations, we substitute the form of a traveling wave for the instant SOA gain coefficient, i.e. $g(z,t) = g(t - z/v_{SOA})$.

In case of TPMZ, the control pulse and the probe clock co-propagate along the SOA at the same speed. Thus the effect of the control pulse on the probe clock seems stationary. Accordingly, the integral for Eq. (5) reduces to the simple form of Eq. (8).

$$h(t) = g(t)L \quad (8)$$

In case of CPMZ, the counter-propagation integral is more complicated in Eq. (9), since the shape of the instant gain coefficient varies with the SOA length when the control pulse and the probe clock travel past each other in the SOA.

$$h(t) = \frac{1}{L}\int_{-L/2}^{L/2} \ln G(t - 2zn_{SOA}/c)dz \quad (9)$$

The intensity transmission characteristics at Port 3 and Port 4 of the XOR gate can be expressed as

$$T_3(t) = \tfrac{1}{4} G_1(t)\{k_1 k_2 + (1-k_1)(1-k_2)R_G \\ - 2\sqrt{k_1 k_2 (1-k_1)(1-k_2)R_G}\, \cos[\Delta\phi_1(t) - \Delta\phi_2(t)]\} \quad (10)$$

$$T_4(t) = \tfrac{1}{4} G_1(t)\{k_1(1-k_2) + (1-k_1)k_2 R_G \\ + 2\sqrt{k_1 k_2 (1-k_1)(1-k_2)R_G}\, \cos[\Delta\phi_1(t) - \Delta\phi_2(t)]\} \quad (11)$$

where $R_G = G_2(t)/G_1(t)$, $k_1$ and $k_2$ are the ratios of coupler C1 and C2, respectively. For simplicity, we set $k_1=k_2=0.5$ in this paper. Thus, the output signal power at Port 3 and 4 can be obtained as

$$P_j(t) = P_{PRB}(t) T_j(t), \quad j = 3,4 \quad (12)$$

where $P_{PRB}(t)$ is the power of the probe signal.

## 4. Simulations and results

To validate the proposal and analyze the gate performance, we conduct numerical simulations, where both input logic signals are 10Gb/s RZ pseudorandom bit sequences (PRBS) with the word length of $2^7-1$. During the fitting procedure with an iterative method, similar in [11], is used to solve such a large space of parameters.

At first, some parameters are treated as constants of representative values for InGaAsP semiconductor materials operating at a wavelength of approximately 1.55 m, as listed in Table 1.

Table 1 Description and values of some fixed parameters in first-step fitting procedure

| Description | Value |
| --- | --- |
| Effective index of SOA | $n_{SOA}$ = 3.62 |
| Nonlinear gain compression factor | $\varepsilon$ = 0.20 W$^{-1}$ |
| General gain factor. | $a_N$ = 2.5×10$^{-20}$ m$^{-3}$ |
| SOA effective active area | $d \cdot W$ = 0.6*1.5 μm$^2$ |
| Contrast ratio of input signal | 30 dB |
| Control pulse center wavelength | $\lambda_{CTRL}$ = 1.546 μm |
| Control pulse-width | $T_{CTRL}$ = 12.5 ps |
| Control pulse energy | $E_{CTRL}$ = 137.50 fJ |
| Probe pulse center wavelength | $\lambda_{PRB}$ = 1.56 μm |
| Probe pulse-width | $T_{PRB}$ = 5 ps |
| Probe pulse energy | $E_{PRB}$ = 3 fJ |

Then, some reasonable guess is made for the scopes of key SOA parameters that exert influence on the gate performance. These parameters include $g_0$, $\tau_C$, $\alpha$, $\Gamma$, $E_{SAT}$, $L$ and so on.

The switching window-width is an important factor to limit the gate operating speed. We study the window-width with various SOA parameters, among which the most effective ones are and $\tau_C$. As shown in Fig. 2, the switching window rises higher under larger $g_0$, but it also extends wider with the increase of $g_0$, which means the output pulses will get sufficient gain, as well as considerable trailing-edge intersymbol interference (ISI). To suppress ISI, a smaller $g_0$ seems more acceptable. It has to be noticed that the switching window will be rather flattened if $g_0$ is too small. Consequently, $g_0$ around 25dB is chosen for 10Gb/s AND operation.

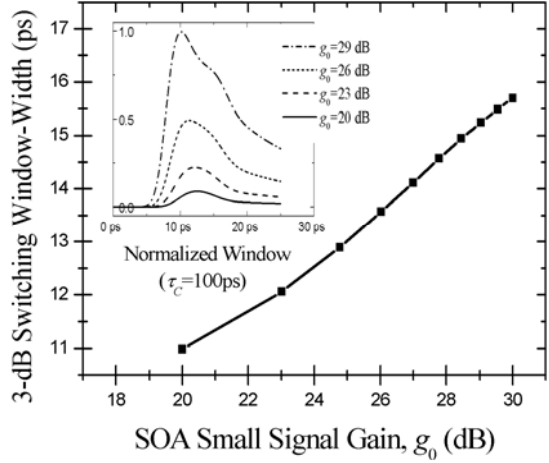

Fig. 2 Evolution of the normalized switching window with small signal gain ($g_0$) for CMZI, $\tau_C$=100ps, $L$=150μm, $\alpha$=3 and $\Gamma$=0.3.

The evolution of the switching window with various $\tau_C$ is plotted in Fig. 3. In general, shorter $\tau_C$ may provide faster SOA response and thus accelerate the gate operation. However, according to Fig. 3 (a), shorter $\tau_C$ introduces serious intrinsic crosstalk due to a nontrivial minor gate window. When $\tau_C$ increases, the minor gate window is lowered apparently and thus the crosstalk is suppressed. Unfortunately, larger $\tau_C$ aggravates the bit-pattern dependence of the gate performance, and thus limit the operation speed, as shown by the uncertainty of the window height in Fig. 3 (b). It seems that a good tradeoff can be achieved when $\tau_C$ is close to the pulse interval. Thus SOAs with $\tau_C$ close to pulses interval are preferable.

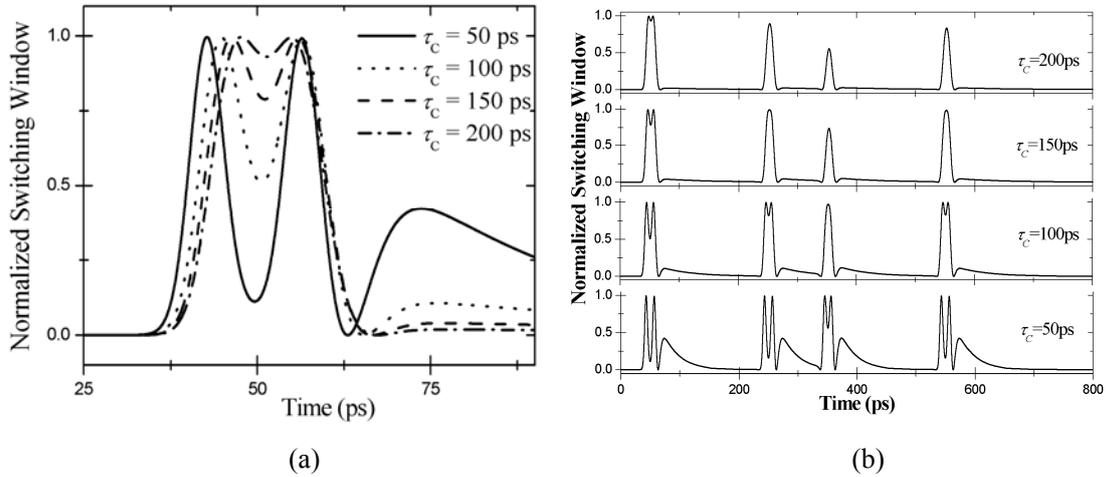

(a)          (b)

Fig. 3 Evolution of switching window with various $\tau_C$, in OPEN state (a) and under PRBS input signals (b), for CPMZ,

$g_0$=25dB, $\Gamma$=0.3, $\alpha$=5.0, $L$=200μm, $E_{SAT}$=2.84pJ, $\Delta\tau$=12.5ps

In accordance with Equ. (7), the larger the α is, the sharper the phase will shift and, therefore, the higher the window will open. However, as α increases, the window becomes distorted and broadened severely, and thus serious ISI deteriorates the gate performance, as shown in Fig. 4 (a). In contrast, the window ceases to split when α decreases, with the advent of strengthened bit-pattern dependence of the AND gate, as shown in Fig. 4 (b). Consequently, SOAs with moderate value of α are favorable.

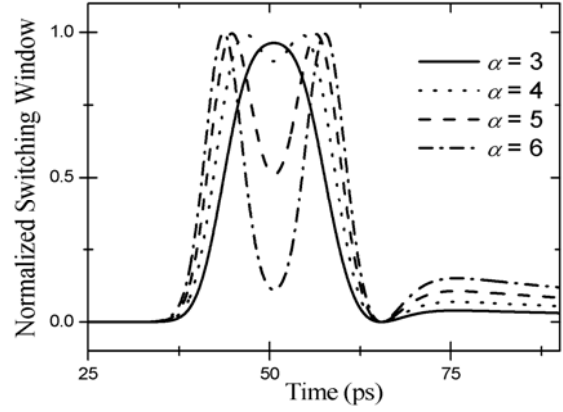

(a)

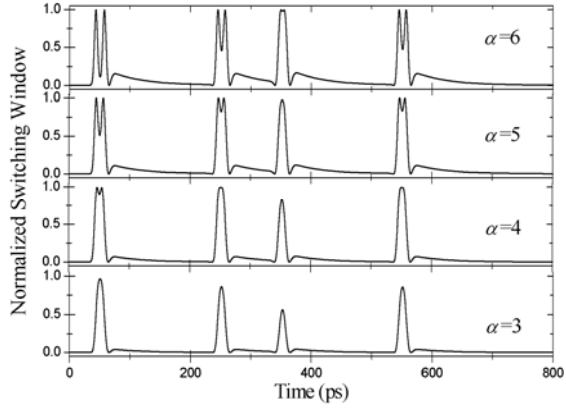

(b)

Fig. 4 Evolution of switching window with various α, in OPEN state (a) and under PRBS input signals (b), for CPMZ,

$g_0$=25dB, $\Gamma$=0.3, $\tau_C$=100ps, $L$=200$\mu$m, $E_{SAT}$=2.84pJ, $\Delta\tau$=12.5ps

Thereafter, setting $g_0$=25dB, $\tau_C$=100ps and $\alpha$=5, we go on study the influence by $\Gamma$, $L$, $\Delta\tau$ and $E_{SAT}$ on CMZI-based XOR gate performance. We select the minimum output contrast ratio (CR) to indicate the opening of the eye diagram and is defined as the ratio of the minimum output peak power for ONE to the maximum output peak power for ZERO in dB.

As shown in Fig. 5, CR degrades sharply when $\Gamma$ increases. It is also captured that CR increases almost linearly with $E_{SAT}$. To keep a sufficient $E_{SAT}$, a large SOA active area (d·W) and a small $\Gamma$ are expected. The knot is that $\Gamma$ usually drops with the increase of d. However, for reason of single mode operation, d should not increase too much and must satisfy $d<\lambda_0/2(n_{SOA}^2-n_2^2)^{1/2}$. To address this problem, many studies have been made, such as the multi-waveguide SOA [14] and assisted light.

In the study of the parameters, it is found that shorter L may yield better gate performance and the SOA length effect on CMZI is obviously stronger than that on TMZI [11]. The parameter, $\Delta\tau$, is also examined and it indicates the gate works better when $\Delta\tau$ is close to the pulse-width.

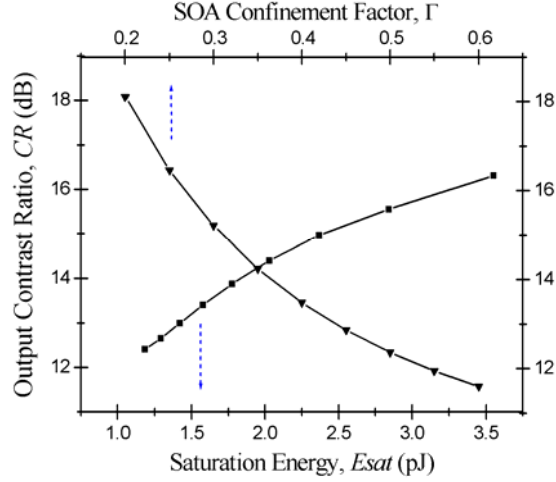

Fig. 5 Output contrast ratio of the AND gate as the functions of $\Gamma$ and $E_{SAT}$

The maximum frequency chirps, defined by $\Delta\nu_{max}(t) = \max\left|\frac{1}{2\pi}\frac{d\phi(t)}{dt}\right|$, under different pulse profiles, are compared in Fig. 6. Generally, the dynamic SOA gain varies more acutely with the increase of the input pulse energy, and thus the absolute value of the chirp augments. For a given pulse energy, the chirp under Super-Gaussian pulses ranks the largest, that under the hyperbolic secant pulses is the second best, while the chirp under Gaussian pulses is the lowest. Therefore, in the following simulations, we select the Super-Gaussian pulse ($m$=3) for a better performance.

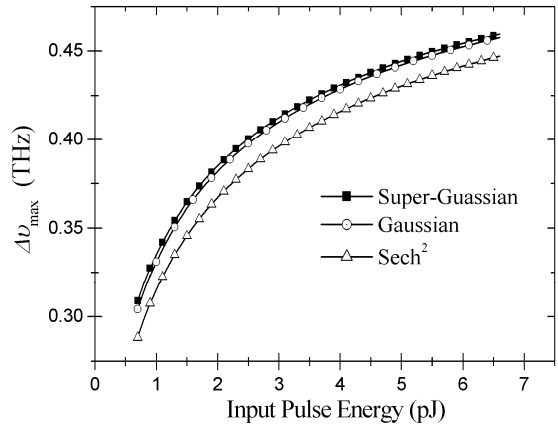

Fig. 6 Maximum chirp of pulses with different profiles and powers

According the fitting procedure above, we select the moderate values of these parameters, as listed in Table 2, and simulate the AND gate operation under 10Gb/s PRBS.

Table 2 Description and values of some fixed parameters in first-step fitting procedure

| Description | Value |
| --- | --- |
| Line-width enhancement factor | $\alpha$=5.0 |
| Carrier lifetime | $\tau_C$=100 ps |
| Length of the SOAs | $L$=200 $\mu$m |
| Saturation energy | $E_{SAT}$=2.84 pJ |
| Time delay | $\Delta\tau$=12.5 ps |
| Small signal gain | $g_0$=25 dB |
| General confinement factor | $\Gamma$=0.39 |

A portion of simulated dynamic gains in both arms of CMZI and the differential phase-shift

($\Delta\phi_1-\Delta\phi_2$) are shown in Fig. 7. The corresponding input and output signals are shown in Fig. 8, where Signal $A$=(01110101)$_2$ and Signal $B$=(10110100), covering all possible Boolean AND operations of two binary bits. The simulations convey that the AND truth-table is realized desirably through SOA-based CMZI with differential inputs. The extinction ratio of the output signal is around 15dB.

As shown in Fig. 7, during each ONE bit-period, the SOA gains toboggans and the probe pulses in both arms undergo abrupt phase changes, while in each ZERO bit-period the gains recover and the phase changes level gradually. Continuous ONE bits in 3$^{rd}$ and 4$^{th}$ periods saturate the SOAs further and continuous ZERO bits help the gains to rise further. That is why $\Delta\phi1-\Delta\phi2$ varies in each bit-period of the simulations. Correspondingly, the maximum $\Delta\phi1-\Delta\phi2$ accounts for the most opened window.

To yield a satisfactory performance, $\Delta\phi_1-\Delta\phi_2$ is unnecessary to be $k\pi$, where $k$ is an odd integer, because the interferometer, which is biased for a null, can switch well even if $|\Delta\phi1-\Delta\phi2|$ is less than $\pi$. Therefore, good switching can be achieved with less pulse energy. Examples are shown in combination of Fig. 7 and 8, where $|\Delta\phi1-\Delta\phi2|<\pi$ for most of the time, while $CR$ still remains over 9dB.

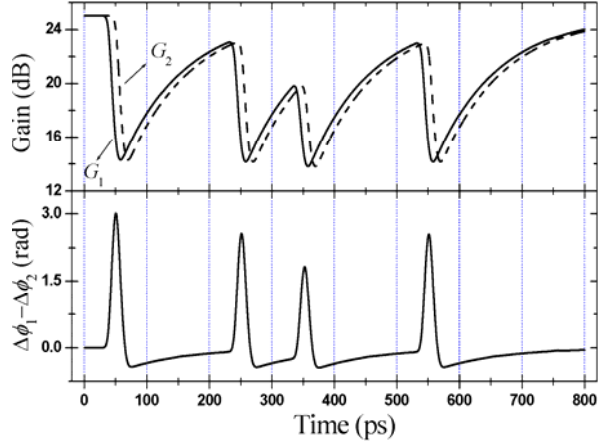

Fig. 7 Dynamic gains in both arms of CMZI ($G_1$ and $G_2$) and differential phase-shift, under 10Gb/s PRBS, $B$=(10110100)$_2$

Figure 2 (a), (b) and (c) show the input pulses ($S_1$, $S_2$) and the output pulses ($S_{AND}$) of our proposed AND gate. We can see that AND operation is realized with an extinction ratio of more than 14dB. Extinction ratio is an important parameter influencing the performance of AND gate. One major factor affecting the extinction ratio is crosstalk, including the intrinsic crosstalk and the channel crosstalk. As illustrated in Figure (c), the intrinsic crosstalk is indicated by the leaking power located at 150 ps when the AND gate should close; while the channel crosstalk appears as the power fluctuation of output signals.

The output peak power uncertainty in Fig. 6 reveals the bit-pattern dependence of the XOR gate, ascribed mainly to the advent of continuous ONE bits. Due to the saturation in SOA and the slow gain recovery, when $A$ and $B$ are identical, the gate window will close to some extent but not completely, which induces inter-symbol crosstalk onto the output signal. An example is shown during the 8$^{th}$ bit-period. Although $A$ and $B$ are both ONE, SOA1 is in deep saturation after continuous ONE bits while SOA2 in relatively shallow saturation. Thus the gate window does not close completely and accounts for the maximum peak power for ZERO bit.

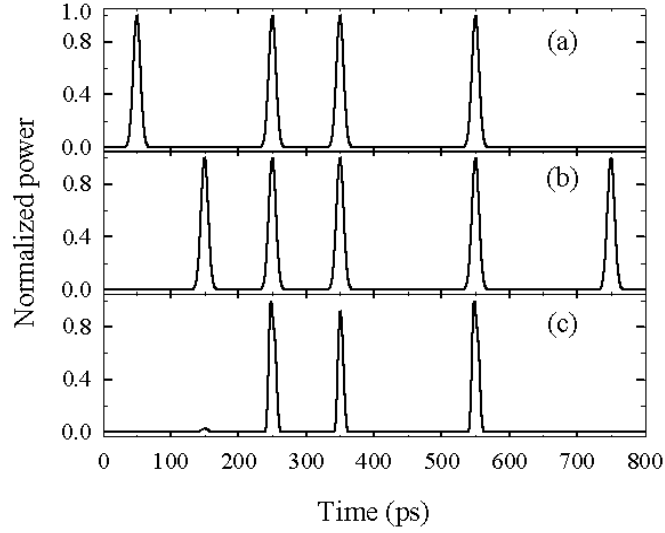

Fig. 8 Simulation results of output signals of AND gate: (a) the input signal $S_2$, serving as control signals; (b) the input signal $S_1$, functioning as probe signals; (c) the output signal of AND operation, $S_{AND}$=15dB

The switching window of Port A and B under the control signal of $S_2$ is shown in Figure 3 (a) and (b), which depends on the bit pattern of $S_2$ due to the XPM effect shown in Figure 3 (c) and (d). According to Figure 2 (a) and Figure 3 (a), the bit ONE of $S_2$ opens Port A, while bit ZERO closes it. Thus, AND operation can be realized with our proposed scheme in an easy and simple way, and the output signal of Port A is the result of AND operation.

In the configuration of AND gate we proposed here, SOA plays an important role. Therefore, it's of great importance to examine it carefully in order to optimize the performance of AND gate. The dynamic gain response of SOA is the underlying factor of the XPM effect and consists of two processes, the fast gain depletion and recovery due to the effect of carrier pulsation which is governed by the carrier lifetime ($\tau_c$). Furthermore, once the dynamic response of SOA is known, the relative phase shift of SOA-MZI depending on the line-width enhancement factor ($\alpha$) of SOA can be determined. In the following, we will examine the effect of the above two parameters of SOA on the performance of AND gate.

Through simulations with various parameters, we find that larger $\tau_c$, $\alpha$ and $\varepsilon$ may slower the falling edge of the window, or result in a minor window right after the major one, and thus introduce serious ISI. Designing a longer SOA length properly may lower the minor window and suppress the crosstalk, however, at the cost of time delay, power penalty and thus the performance degradation.

Assigning moderate values to SOA parameters within reasonable limits, we study the gate performance with respect to the optical signal parameters, such as the input pulse energy and the pulse-width. There is no wonder that shorter pulse-width is preferable to improve the AND operation for both TMZI and CMZI. As revealed in Fig. 8, for 20Gb/s operation, the pulse-width can extends to more than 12ps while the $CR$ over 10dB is maintained. For 40Gb/s operation, when the control or the probe pulse width exceeds 10ps, the gate window will close completely under given conditions. Another phenomenon is that the gate performance begins to degenerate when pulse-width<2ps, for reason of a secondary switching window due to the intra-band process.

Fig. 8 Output contrast ratio vs. control pulse-width for CMZI, with signal rate as a parameter, $g_0$=25dB, $\Gamma$=0.3, $\tau_C$=100ps, $L$=100μm, $\alpha$=3 and $E_{CTRL}$=250fJ

The probe pulse energy should be too small to modify the optical properties of the SOA. Fixing the values of the other parameters, we enhance the control pulse energy ($E_{CTRL}$) gradually and examine the output signals. When $E_{CTRL}$ exceeds a certain value, e.g. 0.4pJ in Fig. 9, the improvement by enhancing $E_{CTRL}$ becomes unapparent.

Fig. 9 Output contrast ratio (*CR*) vs. input control pulse energy for both TMZI and CMZI, with signal rate as a parameter,

$g_0$=25dB, $\Gamma$=0.3, $\tau_C$=100ps, $L$=100μm, $\alpha$=3 and $T_{CTRL}$=5ps.

According to the above discussion, we state some rules regarding the influence by key parameters and the choice of them.

1) Shorter *L* may yield better gate performance; moreover, the length effect on CMZI is obviously stronger than that on TMZI;

2) Larger $g_0$, smaller $\Gamma$, $\varepsilon$, $\tau_X$ and $\alpha$ all lead to a high *CR*; generally, *CR* increases almost linearly with $P_{sat}$;

3) Shorter pulse-width is preferable to improve the OXR operation for both TMZI and CMZI; however, the performance of CMZI gate begins to degenerate when pulse-width<2ps, due to the appearance of a secondary switching window;

4) The improvement by enhancing $E_{CTRL}$ is relatively unapparent;

5) To yield a satisfactory performance, a differential phase shift of $\kappa\pi$ is unnecessary; therefore, good switching can be achieved with less pulse energy than ideally needed;

6) The bit-pattern effect appears more seriously with the advent of continuous incoming ONE bits;

7) Larger $\tau_X$ and small $g_0$ may slower the falling edge of the window, and large $\alpha$ may result in a secondary switching window, all of which will introduce serious trailing-edge ISI, aggravate the pattern dependence and thus limit the operation speed;

8) With values of the other factors fixed, to get a *CR* over 8dB, the parameter scopes for 20Gb/s XOR operation are $L$<250μm, $\tau_c$<150ps, $\Gamma$<0.5, $\alpha$<5 and $P_{sat}$>5pJ; the scopes become tighter for 40Gb/s operation, i.e. $L$<250μm, $\tau_C$<100ps, $\Gamma$<0.35, $\alpha$<4, $P_{sat}$>11pJ, $g_0$>25dB, $E_{CTRL}$>0.15pJ and pulse-width<6ps.

## 5. Conclusions

In this paper, a novel high-speed all-optical AND gate based on SOA-MZI working at 10Gbps is proposed. Theory analysis and numerical simulations are performed to validate our proposal. In addition, the optimum parameters of SOA, such as the carrier lifetime and the line-width enhancement factor, are also investigated. The results show that for AND operation at the speed of 10Gbps, the carrier lifetime and the line-width enhancement factor play an important role in determining the performance of AND gate.

We have analyzed the performance of SOA-MZI-based XOR gate through numerical simulations, in searching of the optimal parameters as well as the suitable operation condition. Several rules and limits on the parameter design have been stated. With properly designed parameters, the high-speed all-optical XOR gate can be realized with fairly high performance using SOA-MZI. Generally, the optimal values of these key parameters are not unique and should be determined in a comprehensive way.

We have to point out that the results of our preliminary work give an approximate solution to the high-speed XOR gate design. The influence on the XOR gate by pulse wavelength is not studied separately in this paper, which is more complicated since the wavelength correlates with many SOA parameters. With consideration of ASE, the scopes of key parameters will be tighter. Finally, if the control pulse energy exceeds 1pJ, the effects of TPA and UNR should be taken into account.